# Experience with multi-threaded C++ applications in the ATLAS DataFlow software

S. Gadomski[1], H.P. Beck, C. Haeberli, V. Perez Reale
*Laboratory for High Energy Physics, University of Bern, Switzerland*

M. Abolins, Y. Ermoline, R. Hauser
*Michigan State University, Department of Physics and Astronomy, East Lansing, Michigan*

A. Dos Anjos, M. Losada Maia
*Universidade Federal do Rio de Janeiro, COPPE/EE, Rio de Janeiro*

M. Barisonzi[2], H. Boterenbrood, P. Jansweijer, G. Kieft, J. Vermeulen
*NIKHEF, Amsterdam*

M. Beretta, M.L. Ferrer, W. Liu
*Laboratori Nazionali di Frascati dell' I.N.FN., Fracasti*

R. Blair, J. Dawson, J. Schlereth
*Argonne National Laboratory, Argonne, Illinois*

J. Bogaerts, M. Ciobotaru, E. Palencia Cortezon, B. DiGirolamo, R. Dobinson, D. Francis,
S. Gameiro, P. Golonka, B. Gorini, M. Gruwe, G. Lehmann, S. Haas, M. Joos, E. Knezo, T. Maeno,
L. Mapelli, B. Martin, R. McLaren, C. Meirosu, G. Mornacchi, I. Papadopoulos, J. Petersen,
P. de Matos Lopes Pinto, D. Prigent, R. Spiwoks, S. Stancu, L. Tremblet, P. Werner,
*CERN, Geneva, Switzerland*

D. Botterill, F. Wickens
*Rutherford Appleton Laboratory, Chilton, Didcot*

R. Cranfield, G. Crone
*Department of Physics and Astronomy, University College London, London*

B. Green, A. Misiejuk, J. Strong
*Department of Physics, Royal Holloway and Bedford New College, University of London, Egham*

Y. Hasegawa
*Department of Physics, Faculty of Science, Shinshu University, Matsumoto*

R. Hughes-Jones
*Department of Physics and Astronomy, University of Manchester, Manchester*

A. Kaczmarska, K. Korcyl, M. Zurek
*Henryk Niewodniczanski Institute of Nuclear Physics, Cracow*

A. Kugel, M. Müller, C. Hinkelbein, M. Yu,
*Lehrstuhl für Informatik V, Universität Mannheim, Mannheim*

A. Lankford, R. Mommsen,
*University of California, Irvine, California*

M. LeVine,
*Brookhaven National Laboratory (BNL), Upton, New York*

Y. Nagasaka,
*Hiroshima Institute of Technology, Hiroshima*

K. Nakayoshi, Y. Yasu,
*KEK, High Energy Accelerator Research Organisation, Tsukuba*

M. Shimojima,
*Department of Electrical Engineering, Nagasaki Institute of Applied Science, Nagasaki*

G. Zobernig,
*Department of Physics, University of Wisconsin, Madison, Wisconsin*

---

[1] on leave from INP Cracow, Poland
[2] also from Universiteit Twente, Enschede, Netherlands

**MOGT011**

The DataFlow is sub-system of the ATLAS data acquisition responsible for the reception, buffering and subsequent movement of partial and full event data to the higher level triggers: Level 2 and Event Filter. The design of the software is based on OO methodology and its implementation relies heavily on the use of posix threads and the Standard Template Library. This article presents our experience with Linux, posix threads and the Standard Template Library in the real time environment of the ATLAS data flow.

## 1. INTRODUCTION

The DataFlow system is a part of ATLAS Trigger/DAQ. The system is responsible for reception, buffering and movement of event data to and from the high level triggers (HLT), known in ATLAS as the Level 2 Trigger (LVL2) and the Event Filter (EF). The system consists of software applications, which run on standard Linux PCs connected to standard Ethernet networks. The overview of the DataFlow project is given in [1]. The networking aspects of the projects are presented in [2]. In this article we describe the experience with the Data Flow software.

The DataFlow software applications are written in C++. The applications have different functions, described in more detail in [1]. All the applications also have to perform certain common tasks such as:
- sending data over network and receiving it,
- accessing configuration data base,
- providing monitoring information,
- executing state transitions,
- reporting errors.

These and other functions are provided by an OO framework, on which all the DataFlow applications are based.

All the applications are implemented using multiple threads of execution. Threads are "light weight processes", which are scheduled separately by the operating system, but share the resources of the executable, in particular the memory. The use of multiple threads enables a more effective use of CPU by DataFlow applications. Data transfers imply latency. While one of the threads waits for data, other threads can use the CPU to perform other tasks.

Several functions of the framework, such as monitoring, are also implemented as separate threads. These "service" threads do not take large fractions of CPU time, but they are ready to take action responding to an external request. Threads can also be activated at regular (and configurable) intervals, performing tasks related to time outs, most notably corrective actions in case of lost messages.

The DataFlow applications need to be able to work with "unsafe" connection-less protocols such as UDP/IP. Using "safe" protocols, such as TCP/IP can penalize the performance of the DataFlow system. In many cases it is better to deal with potential packet loss at the application level, rather then in the network protocol. Robustness against packet loss is required for DataFlow applications.

The Standard Template Library (STL) [3] has become widely used in C++ programs. The library provides commonly used data structures such as vectors, lists and maps. The DataFlow software is using the containers of the STL library.

The current prototype of the DataFlow software was developed in 2001-2002. Since Autumn 2002 the performance of the software is measured systematically. The measurement program has validated the concepts used in the DataFlow software for the Technical Design Review of the ATLAS Trigger and DAQ, which is due in June 2003. The measurements have led to occasional optimizations of the software.

The prototype DataFlow software will also be deployed as the DAQ during beam test of ATLAS detector prototypes starting in May 2003. Preparation of the software for the beam test required several improvements in stability and ease of deployment of the software.

Some lessons learned during the development and testing of DataFlow software are described in Section 2. We offer our conclusions in Section 3.

## 2. EXPERIENCE

In this section we present problems, which were encountered during development and testing of the DataFlow software, as well as the adopted solutions. In order to explain the problems we briefly present the use of threads in selected DataFlow applications.

### 2.1. STL Containers in multi-threaded applications

The problem with using STL containers in DataFlow applications was first observed during performance measurements of the second level trigger Processing Unit (L2PU). The L2PU application will run the algorithms that will access event data and produce the LVL2 decision on each ATLAS event. It is expected that there will be several hundred PCs running as L2PUs in the ATLAS Trigger/DAQ system.

An L2PU has a configurable number of "worker threads". Each worker thread processes one event at a time, asking for event data when the physics selection algorithm needs it.

Having several worker threads compensates for the latency of obtaining event data. While one worker thread waits for data, other threads can use the CPU to process their events. A schematic diagram of the L2PU threads is shown in Fig.1.

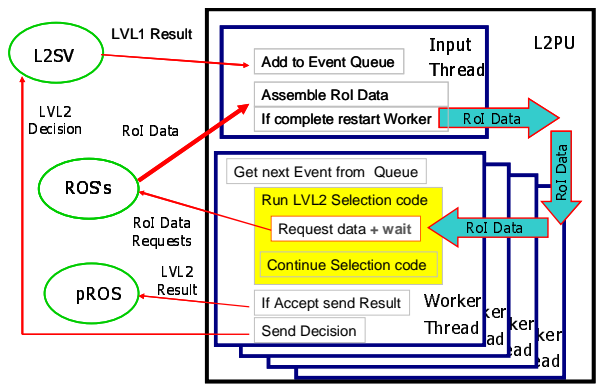

Fig.1 Use of threads in the second level trigger processing unit (L2PU). Multiple Worker Threads run event selection algorithms and request the event data from ROSes.

During tests of L2PU it was discovered that increasing the number of threads did not improve the performance as it was expected. The worker threads did not seem to run truly independently of each other. A diagnostic tool called Visual Threads [4] was helpful in diagnosing the problem. The tool uses instrumented threads library. It enables to trace the switching of context and to understand when threads can block each other. The analysis done on the L2PU has shown that threads were blocked when accessing STL containers. By default all the containers, which are created and used in different threads, share one common memory pool. Simultaneous access to the same memory by different threads could lead to corruption. The STL protects against that by "mutex locks". As the name suggests a mutex lock provides mutual exclusion. Only one thread can access the memory pool at any given time. This can cause a performance penalty because threads can be blocked when accessing STL containers. Because the memory pool is global per executable, even the most local STL containers, which are created and used in one thread, can activate the lock and cause contention between threads.

The problem can be avoided by having separate memory pools for STL containers used in different threads. When an STL container is declared it is possible to choose the memory allocator used by the library.

With some compilers it is also possible to change the allocator globally, for all the containers at once, using a compiler flag. The latter solution could be quickly applied to the L2PU application in order to verify the hypothesis presented here. Using the "pthread" allocator, applied globally to all containers, brought a significant improvement of performance in the L2PU. The speed of the application could be improved by as much as a factor of four under some conditions.

Using the memory pool of STL allocated per thread can lead to a problem when STL containers are created in one thread and deleted in another one. This way the amount of available memory is constantly shifted from one thread to another, which may not be sustainable in the long run.

The most practical solution to this problem is to use a special memory allocator, which can track the "migration" of memory between threads. An allocator like that can take corrective actions, allocating memory in one thread and freeing it in another. For performance reasons this corrective action is undertaken once in a while, i.e. not every time an STL container is created or deleted. This allows an effective use of the STL containers at high rates.

The solution with a dedicated allocator is now being implemented in the DataFlow software. All the STL containers need to be revisited. Declarations of the containers that are used at high rates need to be changed. However, it is already understood that a solution exists. The STL containers can be used in multi-threaded applications without causing loss of performance.

### 2.2. Controlling the network traffic

Sub Farm Interface (SFI) is the DataFlow application responsible for event building in ATLAS. An event accepted by the second level trigger is assigned to an SFI. The SFI is given the LVL1 ID of the event. The task of the SFI is to ask for data from a number of sources and to build a complete event. Depending on an option chosen by ATLAS the SFI may receive data from around 200 or from around 1600 sources connected to the network (see [1] for more details). Once the SFI has a complete event, the event data can be deleted from Readout Buffers. The event can be kept by the SFI and can be handed over to the Event Filter farm for further processing.

In order to perform its tasks effectively the SFI has separate threads for the following tasks:
  - requesting data from the data sources,
  - receiving the data,
  - assembling events from fragments,
  - sending events to the event filter.

The four mentioned threads all need to work at a high rate. These threads are shown schematically in Fig.2.

In addition the SFI has other threads, which do not operate at high rate, and which take care of such aspects as operational monitoring, monitoring of the event data or initiating corrective action if data is missing for some unfinished events.

The design of SFI is a result of optimizations. It was driven by performance measurements of a fully functional prototype performing all the necessary I/O. The environment on the measurement test bed was identical to that in which the SFI will be working in ATLAS. The network transaction needed to obtain data and to ship it out were like expected in the full system.

The SFI can reach the optimum performance because it controls the flow of messages to which it is exposed. Data fragments from hundreds of sources have to arrive to the same input of an SFI. This brings the risk of collision.

**MOGT011**

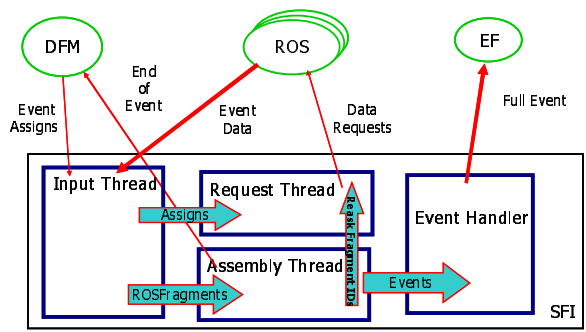

Fig. 2 Use of threads in the sub-farm interface (SFI). Delegating work from the Input Thread to the others improves performance.

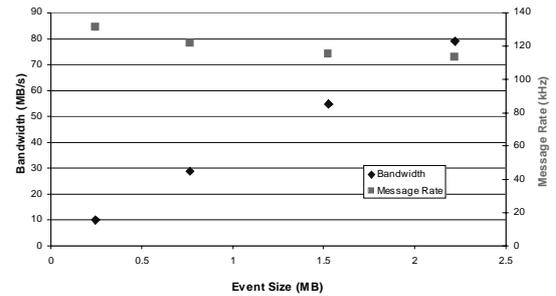

Fig.3 Performance of the SFI. Bandwidth and total message rate (incoming + outgoing) are shown as a function of the event size.

The messages containing event fragments can be lost in queues at the ports of switches. It was also observed that messages can be lost in the kernel buffer of the destination computer if the application can not read them fast enough.

The SFI is robust against packet loss. Missing fragments of events are asked for again. However re-asking causes performance loss and should be avoided. The only effective way to achieve this is by controlling the flow in the SFI itself. The SFI limits the number of requests for data that are outstanding at any given time. This gives an automatic adjustment of data rate, compensating for all possible limitations of bandwidth in the network. Thanks to this mechanism the SFI can receive data at a rate that is a large fraction of the network line speed.

Other experience accumulated during the development and optimization of the SFI was multi-fold. The observations related to STL containers, described in the previous section, were confirmed with the SFI application. More improvements were obtained by avoiding:
- system calls,
- creations of objects,
- contention of threads related to sharing objects.

Another improvement was reached by reducing the frequency of thread switching. The thread that assembles event fragments, as well as the thread requesting data, have outstanding work which depends on the incoming fragments. It is better to activate them less often and to let them process more fragments (or send more requests) at a time. This issue was not predicted in advance, but a significant performance gain of 14% was reached by reducing the frequency of thread switching.

Fig.3 shows the performance of the SFI when doing input only. The size of the events was varied. The messages containing event data were limited in size to a single Ethernet frame, around 1.4 kB. The total message rate (outgoing requests for data + icoming fragments) was reaching 130 kHz. With full frames the data was collected at 79 MB/s.

When data sources send multi-frame messages the SFI can sustain input data rate up to 95 MB/s, which is 76% of the bandwidth of the gigabit Ethernet. When doing simultaneous input and output the SFI can reach the speed of 70 MB/s. The performance is limited by the speed of the CPU. The results presented here were obtained on a 2.4 GHz PC.

The optimization of the complete application done in a realistic environment has enabled to reach a performance sufficient for ATLAS data flow.

## 2.3. Scheduling of threads in Linux

A problem with thread scheduling arose in the Readout System (ROS) application. The ROS receives requests for data from the L2PU and from the SFI. In response the ROS collects data from readout buffers (ROB) and sends it to the requesting process over network.

Both the ROS and the SFI collect data from several sources. The requirement on the total data rate is the same. However other requirements are very different between the ROS and the SFI, as illustrated in table 1.

| Requirement | ROS | SFI |
| --- | --- | --- |
| Request Rate | 24 kHz LVL2<br>3 kHz EB | 50 Hz |
| Data to send per request | 2 kB LVL2<br>8 kB EB | 1.5 MB |
| Number of data sources per request | ≈2 (LVL2 average)<br>12 EB | 200 to 1600 |
| Total data rate | 72 MB/s | 75 MB/s |

Table 1. Requirements on ROS and SFI in the ATLAS DataFlow system. For the ROS the requirements from the second level trigger (data transfers to L2PUs) and from the Event Building (data transfers to SFIs) are shown separately.

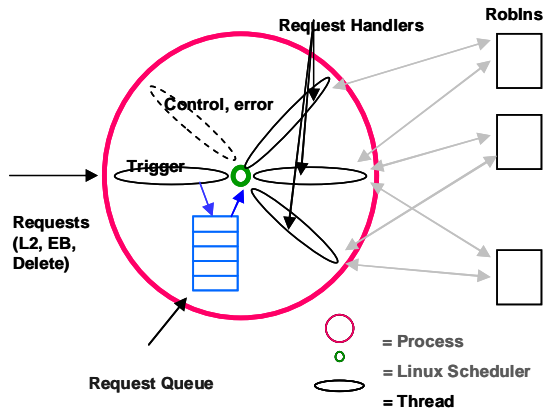

Fig.4. Use of threads in the Readout System (ROS). Multiple Request Handler threads obtain data from RobIns.

The use of threads in the ROS is different with respect to the SFI, resembling more the L2PU. There is a configurable number of request handlers, each serving one request for data at a time. The threads of the ROS are shown schematically in Fig.4.

A request thread needs to take action after data from ROBs appears in memory buffers. As the ROS is designed to work without interrupts, the only way a threads can know that data is ready is by checking a memory location. In order not to block the CPU by polling in a loop, a `yield()` instruction is used. This instruction returns control from a thread. A problem with this approach was discovered during testing of the ROS. It was expected that latency of obtaining data, which was simulated in the measurements, can be compensated by increasing the number of worker threads.

For a sufficiently large number of threads one of them would always have data available and the CPU would never be idle. Increasing the number of request threads did not bring the expected increase of performance, as one can observe in Fig.5 (lower curve).

An investigation uncovered that the operating system was switching only between some worker threads, not giving context to other threads for a while. It was understood that a thread in which `yield()` was called could be put on hold for a time up to 100 ms. The thread scheduling algorithm was reset with the frequency given by the HZ parameter, by default 100 Hz.

A remedy is to apply a kernel patch. For CERN Red Hat Linux 7.3 there exists an official patch that changes the scheduling policy [5]. After a patch like that is applied the scheduling is done in a round-robin fashion, not excluding threads that have called `yield()`. This gives the expected performance of the ROS. The improvement of performance introduced by the patch is shown in Fig.5. The ROS running on a PC with the patched kernel meets the requirements of the ATLAS DataFlow system

## 3. CONCLUSIONS

The DataFlow subsystem of the ATLAS DAQ has a prototype implementation based on multi threaded C++ programs running on Linux PCs and exchanging data via gigabit Ethernet network. The performance of the prototype will be documented in the Technical Design Report published in June 2003.

During the development and testing of the prototype applications several technical problems, mostly related to having many threads, were discovered, understood and solved. The DataFlow system based on the chosen approach can meet performance requirements of the ATLAS experiment.

### Acknowledgments

We wish to thank the ATLAS TDAQ Online Software group for providing a system and useful tools to control, configure and operate large-scale distributed testbed setups.

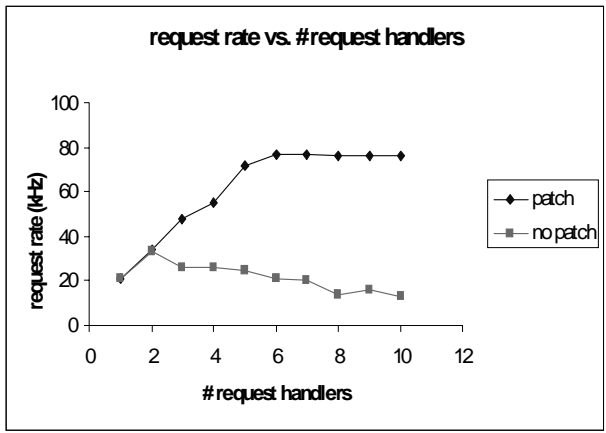

Fig. 5. Performance of the ROS with and without the kernel patch. The performance is measured by the request rate that can be sustained. With the kernel patch the ROS can compensate for latency of obtaining data by having more request handlers.

**MOGT011**

**MOGT011**